# Quantitative Risk Assessment in Radiation Oncology via LLM-Powered Root Cause Analysis of Incident Reports

Yuntao Wang, MS[1], Siamak P. Najad-Davarani, PhD[1], Elizabeth Bossart, PhD[1], Matthew T. Studenski, PhD[1], Mariluz De Ornelas, PhD[1], Yunze Yang, PhD[1*]

[1]Department of Radiation Oncology, University of Miami, FL 33136, USA

*Corresponding Author

Yunze Yang, PhD, Assistant Professor of Radiation Oncology

Department of Radiation Oncology, Sylvester Comprehensive Cancer Center, the University of Miami, 1475 NW 12th Ave, Miami, FL 33136

E-mail: yxy971@med.miami.edu

Author Responsible for Statistical Analysis Name & Email Address

Yunze Yang, PhD, yxy971@med.miami.edu

**Abstract**

Background

Modern large language models (LLMs) offer powerful reasoning that converts narratives into structured, taxonomy-aligned data, revealing patterns across planning, delivery, and verification. Embedded as agentic tools, LLMs can assist root-cause analysis and risk assessment (e.g., failure mode and effect analysis FMEA), produce auditable rationales, and draft targeted mitigation actions.

Methods

We developed a data-driven pipeline utilizing an LLM to perform automated root cause analysis on 254 institutional safety incidents. The LLM systematically classified each incident into structured taxonomies for radiotherapy pathway steps and contributory factors. Subsequent quantitative analyses included descriptive statistics, Analysis of Variance (ANOVA), multiple Ordinal Logistic Regression (OLR) analyses to identify predictors of event severity, and Association Rule Mining (ARM) to uncover systemic vulnerabilities.

Results

The high-level Ordinal Logistic Regression (OLR) models identified specific, significant drivers of severity. The Pathway model was statistically significant (Pseudo $R^2$ = 0.033, LR p = 0.015), as was the Responsibility model (Pseudo $R^2$ = 0.028, LR p < 0.001). Association Rule Mining (ARM) identified high-confidence systemic rules, such as "CF5 Teamwork, management and organisational" (n = 8, Conf = 1.0) and the high-frequency link between "(11) Pre-treatment planning process" and "CF2 Procedural" (n = 152, Conf = 0.916).

## Conclusion

The LLM-powered, data-driven framework provides a more objective and powerful methodology for risk assessment than traditional approaches. Our findings empirically demonstrate that interventions focused on fortifying high-risk process steps and mitigating systemic failures are most effective for improving patient safety.

## Introduction

Patient safety is paramount in radiation oncology, a field where technical complexity and intricate clinical workflows create a high-risk environment necessitating rigorous error prevention[1]. Continuous Quality Improvement (CQI) programs are rigorously implemented in healthcare, which provide a structured framework for embedding risk management into clinical operations to systematically enhance patient safety. Within this paradigm, prospective methodologies such as Failure Mode and Effects Analysis (FMEA) are leveraged to proactively identify and mitigate latent vulnerabilities within complex processes before they result in adverse events[2]. Conversely, retrospective tools like Root Cause Analysis (RCA) are employed to investigate incidents, focusing on the discovery of underlying system failures rather than on individual culpability. The integration of these proactive and reactive analyses creates an essential feedback loop, allowing for iterative process refinement and the continuous improvement of safety and quality in patient care[3,4].

In FMEA, a multidisciplinary team maps a clinical process, brainstorms potential failure modes, and assigns semi-quantitative scores for Severity (S), Occurrence (O), and Detectability (D) to calculate a Risk Priority Number ($RPN = S \times O \times D$), which guides the prioritization of mitigation efforts[5]. The American Association of Physicists in Medicine (AAPM) Task Group 100 (TG-100) report formally established the framework for applying such risk-based analysis techniques to quality management in radiation therapy.

Despite its widespread adoption, traditional FMEA is constrained by its inherent subjectivity[6-8]. A significant body of literature has reviewed the limitations of FMEA, noting its reliance on the clinical judgment, experience, and imagination of the expert team rather than empirical data[6].

This reliance on a priori knowledge can introduce significant uncertainties and biases, potentially causing critical failure modes to be overlooked or their true frequency misjudged, thereby misdirecting valuable quality improvement resources.

In contrast, incident learning systems (ILS), such as the Radiation Oncology Incident Learning System (RO-ILS), provide a rich source of retrospective, empirical data on real-world safety events[9,10]. Each report, containing detailed narratives of errors and near-misses, represents an objective observation of a failure within the clinical workflow, offering a direct window into a system's true vulnerabilities[11,12]. Analyses of national ILS data and institutional systems have been instrumental in identifying common error pathways and informing safety improvements, demonstrating significant reductions in errors through systematic learning[10,11,13,14].

However, the primary barrier to leveraging the full potential of ILS data has been the unstructured, free-text format of the reports, which is difficult to analyze systematically and at scale[15]. Manual review is resource-intensive and subject to inter-reviewer variability, creating a significant bottleneck that has limited the ability to perform robust, quantitative risk assessments based on this empirical data.

The recent and rapid advancement of Large Language Models (LLMs) presents a transformative opportunity to overcome this long-standing challenge. LLMs are a form of artificial intelligence (AI) designed to understand, interpret, and generate human language with remarkable acuity, making them particularly well-suited for automating tasks that previously required human expertise, such as Root Cause Analysis (RCA) of safety incidents[16]. Foundational work has demonstrated the efficacy of LLMs in categorizing patient safety reports and analyzing clinical text[17], including the use of statistical modeling for semi-automated topic identification of radiation oncology reports[18]. Recent studies have specifically shown promise in applying these techniques to radiation oncology incidents, such as for automated triaging[19], automated error labeling[20], and full root cause analysis[16].

This study aims to develop and validate a novel, data-driven framework for quantitative risk assessment in radiation oncology. This framework leverages the power of an LLM to automate the analysis of institutional incident reports, transforming unstructured narratives into a standardized, structured database. By applying a suite of quantitative methods, including descriptive statistics, Analysis of Variance (ANOVA), Ordinal Logistic Regression (OLR), and Association Rule Mining (ARM), to this newly structured data, this work seeks to move beyond subjective risk assessment. The objective is to demonstrate a methodology that can empirically identify the most significant predictors of event severity and uncover the latent network of systemic vulnerabilities within a clinical radiation oncology practice, thereby enabling truly evidence-based safety interventions.

## Materials and Methods

### RO-ILS Incident Reports

The data for this study consisted of incident reports from RO-ILS collected at the University of Miami, Sylvester Comprehensive Cancer Center, between October 2022 and June 2025. These submitted reports encompassed a range of safety events, including documented errors, near misses, and unsafe conditions. After initial review and preprocessing, a total of 254 reports were deemed suitable and included in the final analysis.

### Standardized Taxonomies

To ensure objective and reproducible analysis, a structured framework was required to convert the unstructured, narrative-based event descriptions into a quantitative format. This was achieved by adopting two internationally recognized standards. For event classification, we utilized the National Patient Safety Radiotherapy Event Taxonomy published by the UK Health Security Agency[21]. This taxonomy is aligned with the consensus from AAPM[14]. For severity assessment, we referenced the risk analysis framework outlined in the AAPM TG-100 report[5]. Together, these documents provided the definitive conceptual dictionary for the subsequent automated classification.

The application of these taxonomies yielded the primary variables for our analysis:

Pathway Taxonomy: This variable identifies the specific process step within the radiotherapy workflow where an event originated. Following the official UK documentation, our analysis utilized both the 21 major process categories (e.g., 4 Referral for treatment, 11 Pretreatment planning process) and the 213 granular sub-codes (e.g., (11i) Target and organ at risk delineation) to enable analysis at multiple levels of detail.

Contributory Factor (CF) Taxonomy: This variable specifies the root causes of an event. The official taxonomy is structured into five major categories (CF1 Individual, CF2 Procedural, CF3 Technical, CF4 Patient Related, CF5 Teamwork/Management/Organisational) and their respective sub-categories (e.g., CF2d Process design, CF5a Inadequate leadership).

Event Severity: Event severity was quantified using a two-tiered approach based on the TG-100 framework. First, a qualitative Severity Term (e.g., "Wrong dose distribution") and its corresponding numerical range, the S value (e.g., 5-8), were identified. Second, a final integer Severity score (e.g., 7) was assigned to the event. This final numerical Severity score, ranging from 1 to 10, served as the dependent variable in our regression analyses.

**LLM-Automated Root Cause Database Generation**

A systematic pipeline was developed to apply these taxonomies to each incident report using a LLM (Gemini 2.5 Pro., Google). The process involved several distinct steps for each event narrative:

Severity Generation: The LLM was first prompted to analyze the event description in the context of the TG-100 standards. It defined a relevant "severity item" and its corresponding "S value" based on the TG-100 framework. Subsequently, the LLM assigned a final numerical severity score within the range of the determined S value.

Event Classification: Following the severity assignment, the LLM was instructed to classify the event according to the UK National Patient Safety Radiotherapy Event Taxonomy. It strictly adhered to the original classification document to generate four distinct outputs for each event: Classification taxonomy, Pathway subcode taxonomy, Contributory factor taxonomy, and Modality taxonomy. To ensure a comprehensive dataset for subsequent analysis, the model was prompted to generate multiple Pathway subcode and Contributory factor entries where reasonably applicable.

Responsibility Assignment: Finally, based on the generated Contributory factor taxonomies and the original event description, the LLM was prompted to freely assign at least one responsible party (e.g., Physician, Dosimetrist, Department Leadership) corresponding to each identified contributory factor.

This automated process transformed the raw, narrative-based incident reports into a structured, multi-faceted database suitable for quantitative analysis.

**Ordinal Logistic Regression (OLR) of Severity**

We employed Ordinal Logistic Regression (OLR) to analyze the drivers of severity. The ordinal Severity score (1–10) was used as the dependent variable. We conducted five separate OLR analyses, each assessing a distinct set of independent variables, which were coded as binary predictors (i.e., presence or absence): Pathway Taxonomy (major categories); Pathway Taxonomy (sub-categories); Contributory Factor Taxonomy (major categories); Contributory Factor Taxonomy (sub-categories) and Responsible Parties.

The overall fit and statistical significance of each model were evaluated using the McFadden's Pseudo R-squared value and the Likelihood Ratio (LR) p-value. For individual predictors, the log-odds coefficients ($\beta$) were converted into Odds Ratios (OR) to provide an interpretable measure of effect size. An OR greater than 1 indicates increased odds of a higher severity rating, while an OR less than 1 indicates decreased odds. The statistical significance of individual predictors was determined using a p-value ($P > |z|$) threshold of 0.05. These analyses were performed using the “statsmodels” library.

## Systemic Vulnerability Analysis

We utilized Association Rule Mining (ARM) to discover latent co-occurrence patterns and systemic vulnerabilities within the dataset. This technique identifies frequent if-then patterns, expressed as rules in the form of "Antecedent" -> "Consequent". The strength and significance of each discovered rule were quantified using three standard metrics: 1) Support: The proportion of events in the dataset containing both the antecedent and the consequent, indicating the rule's prevalence. It is calculated as $Support(A \rightarrow C) = P(A \cap C)$. 2) Confidence: The conditional probability of the consequent occurring given the antecedent has occurred, measuring the rule's reliability. It is calculated as $Confidence(A \rightarrow C) = P(A \cap C) / P(A)$. 3) Lift: The ratio of the observed confidence to the expected confidence if the antecedent and consequent were independent. A lift value greater than 1 indicates a positive association. It is calculated as $Lift(A \rightarrow C) = Confidence(A \rightarrow C) / P(C)$.

A series of ARM analyses were performed to explore the pairwise relationships between the five sets of variables used in the OLR analyses, with the goal of identifying systemic links between pathway steps, contributory factors, and responsible parties. This analysis was conducted using the "mlxtend" library.

## Statistical Analysis

To formally test for statistical independence between the primary event categories (Pathway, Contributory Factor, and Responsibility), Pearson's chi-square tests were conducted. This analysis was performed on all pairwise combinations of the taxonomies (at both the major and sub-category levels) to determine if observed associations were statistically significant. The analysis was conducted using the "chi2_contingency" function from the "scipy.stats" library.

To conduct a preliminary analysis of severity, Analysis of Variance (ANOVA) was used. All statistical analyses were performed using Python (version 3.13) and its associated scientific libraries, with a p-value of $\leq 0.05$ considered statistically significant for all tests.

## Results

### Characteristics of the Incident Cohort

The application of the LLM-powered pipeline to the 254 incident reports resulted in a richly structured dataset. The model identified 956 Pathway subcode instances, which were aggregated to 609 major pathway category occurrences (counting each major category only once per event). Similarly, 1017 individual Contributory Factor instances were identified, corresponding to 643 major category occurrences. Based on these factors, 686 unique responsible party role assignments were made.

A descriptive analysis of the major categories reveals the primary areas of vulnerability (Table 1). The "Pre-treatment planning process" was the most frequently implicated stage, involved in 166 events. Among root causes, "Individual" and "Procedural" factors were most common, cited in 212 events, with "Slips and lapses" (n = 119) and "Communication" failures (n = 97) being the most prevalent sub-categories. "Department Leadership" was the most frequently assigned responsible party (n = 194).

Table 1 Frequency and Severity of Pathway, Contributory Factor, and Responsibility Categories

| Category Type | Category Name | Event Count (n) | Mean Severity | Median Severity |
|---|---|---|---|---|
| Pathway | Pre-treatment planning process | 166 | 5.663 | 4 |
| | Treatment unit process | 108 | 5.898 | 6 |
| | Communication of intent | 59 | 6.120 | 6 |
| | Booking and administrative process | 55 | 4.236 | 3 |
| | Referral for radiotherapy treatment | 40 | 5.750 | 4 |
| | Pre-treatment activities, imaging | 40 | 6.150 | 6 |
| | On-treatment review process | 23 | 7.174 | 7 |
| | Processes prior to first patient appointment | 18 | 4.944 | 3 |
| | Pre-treatment: patient preparation | 15 | 6.333 | 6 |
| | Staff management | 15 | 5.667 | 6 |
| | End of treatment process | 12 | 7.333 | 7 |
| | Mould room activities | 11 | 5.818 | 6 |
| | Treatment data entry/preparation | 10 | 7.100 | 7 |
| | Infrastructure, strategic and organisational | 10 | 4.800 | 4 |
| | Research and document management | 8 | 8.375 | 10 |
| | Room design, environment and ergonomics | 7 | 6.714 | 7 |
| | Routine machine quality assurance | 6 | 4.833 | 5 |
| | Brachytherapy | 2 | 6.5 | 6.5 |
| | New equipment, techniques and technology | 2 | 6.5 | 6.5 |
| | Timing | 2 | 3 | 3 |
| Contributory Factor | Individual factors | 212 | 5.948 | 6 |
| | Procedural | 212 | 5.656 | 5 |
| | Teamwork, management and organisational | 158 | 5.627 | 4 |
| | Technical | 40 | 5.550 | 5 |
| | Patient related | 17 | 5.529 | 5 |
| | Environmental | 4 | 5.250 | 4.5 |
| Responsibility | Department Leadership | 194 | 5.464 | 4 |
| | Physician | 166 | 5.867 | 6 |
| | Physicist | 80 | 6.438 | 7 |
| | Dosimetrist | 74 | 6.351 | 7 |
| | Therapist | 69 | 6.855 | 7 |
| | IT/Vendor | 38 | 5.421 | 5 |
| | Nurse | 30 | 5.267 | 4 |
| | Scheduler/Admin | 22 | 4.955 | 3.5 |
| | Patient | 13 | 5.462 | 5 |

**Risk Profiles**

To further explore the distributions, relationships, and systemic patterns within the data, a series of visualizations were generated for the five primary variable sets: Pathway Taxonomy, Contributory Factor Taxonomy, and Responsible Parties.

The frequency distributions for each major category are illustrated through pie charts, while Pareto charts are used to visualize the frequencies of the more numerous sub-categories (Figure 1). These charts demonstrate that a disproportionately large number of incidents are associated with a relatively small number of specific process steps and root causes. This highlights the key targets for quality improvement initiatives, such as the most common categories of "Pre-treatment planning process" and "Individual factors".

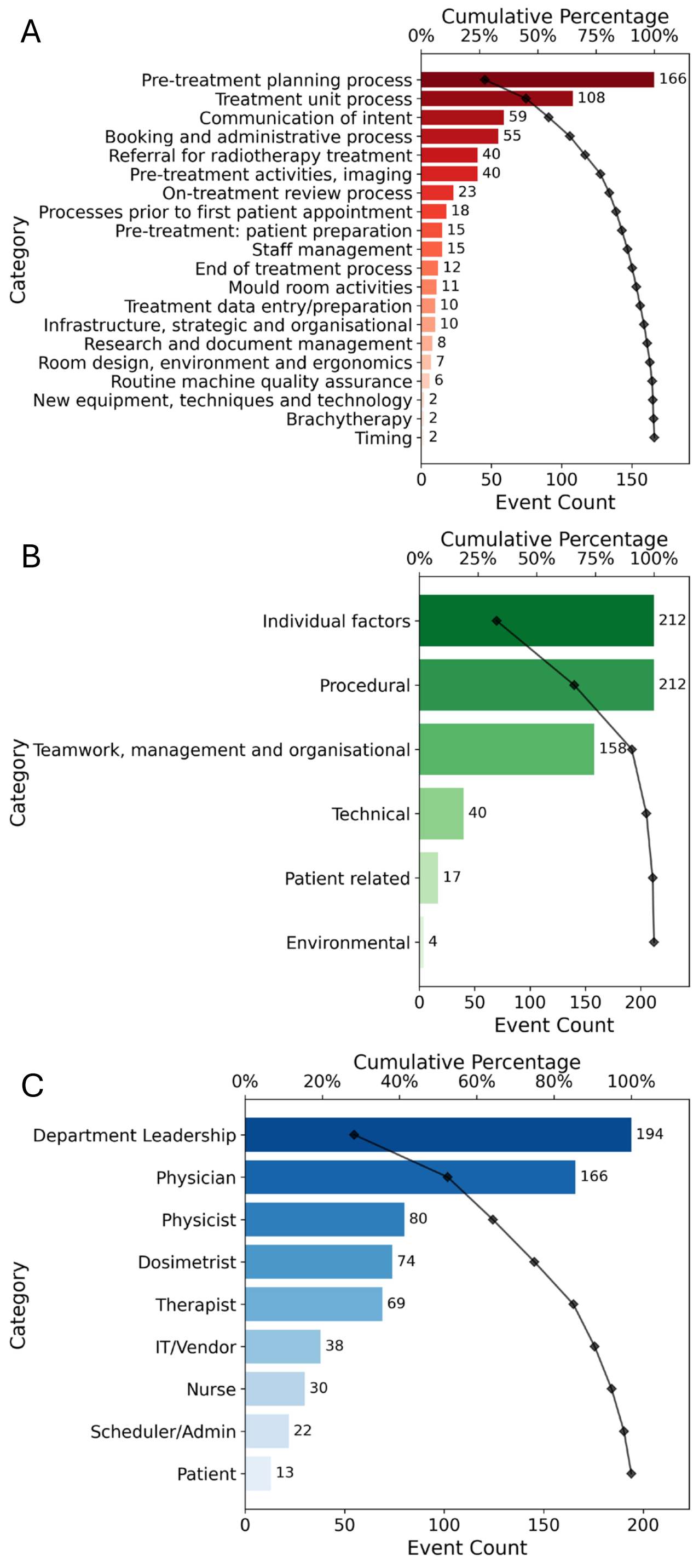

Figure 1 Frequency distribution charts for (A) Pathway Categories, (B) Contributory Factor Categories, (C) Responsibility.

To investigate the relationship between each category and potential harm, the distribution of Severity scores is visualized using box plots (Figure 2). These plots provide a comparative view of the median severity and interquartile range for each category, revealing which types of failures are associated with higher-severity outcomes. For instance, events involving Dosimetrists show a higher median severity compared to other roles.

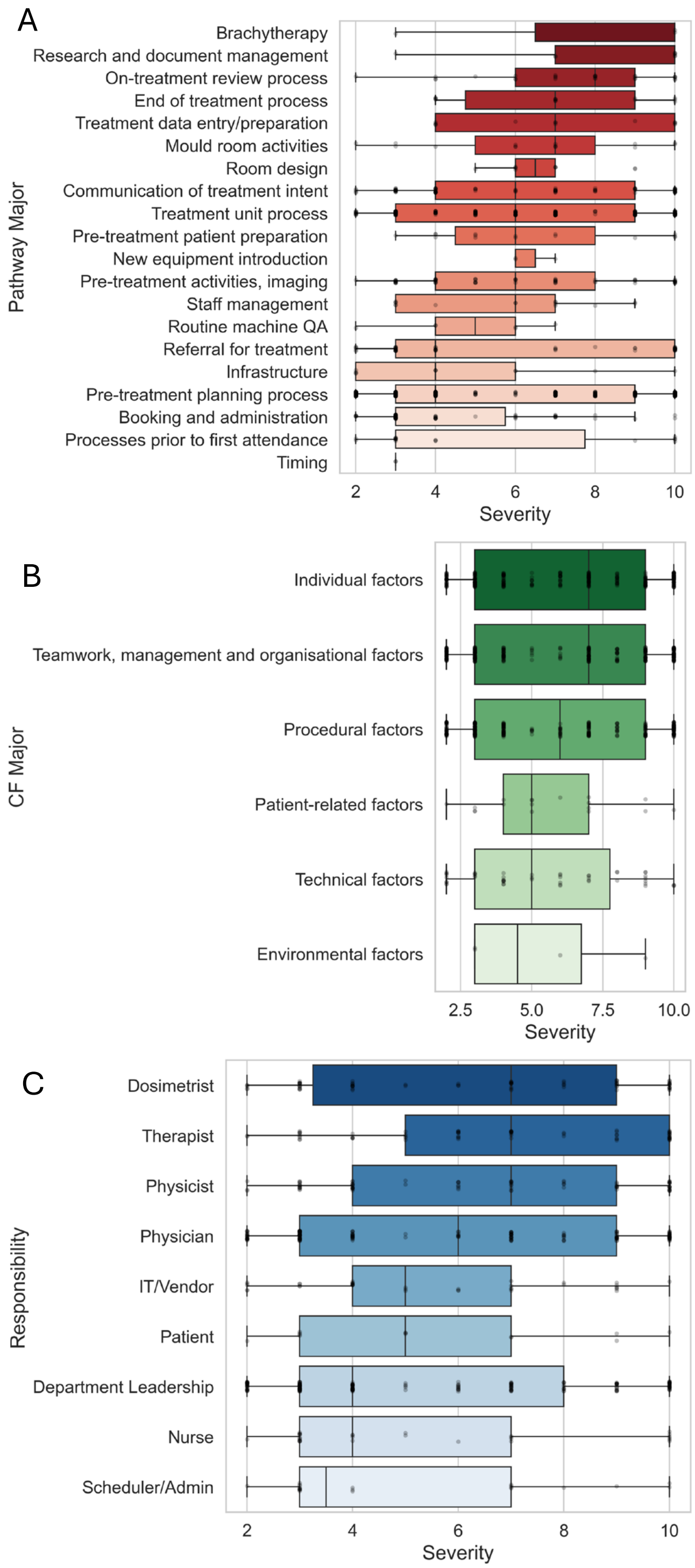

Figure 2 Box Plots of Severity Scores for (A) Pathway Categories, (B) Contributory Factor, (C) Responsibility.

We also examined the co-occurrence patterns between pairs of variable sets using heatmaps (Figure 3). The color intensity in each cell corresponds to the frequency of co-occurrence, highlighting critical intersections of process failures, their root causes, and the involved parties. We identify systemic relationships, such as the strong association between “Procedural” factors and the “Pre-treatment planning process”. “Department Leadership” shows the highest co-occurrences to all Pathways and Contributing Factors.

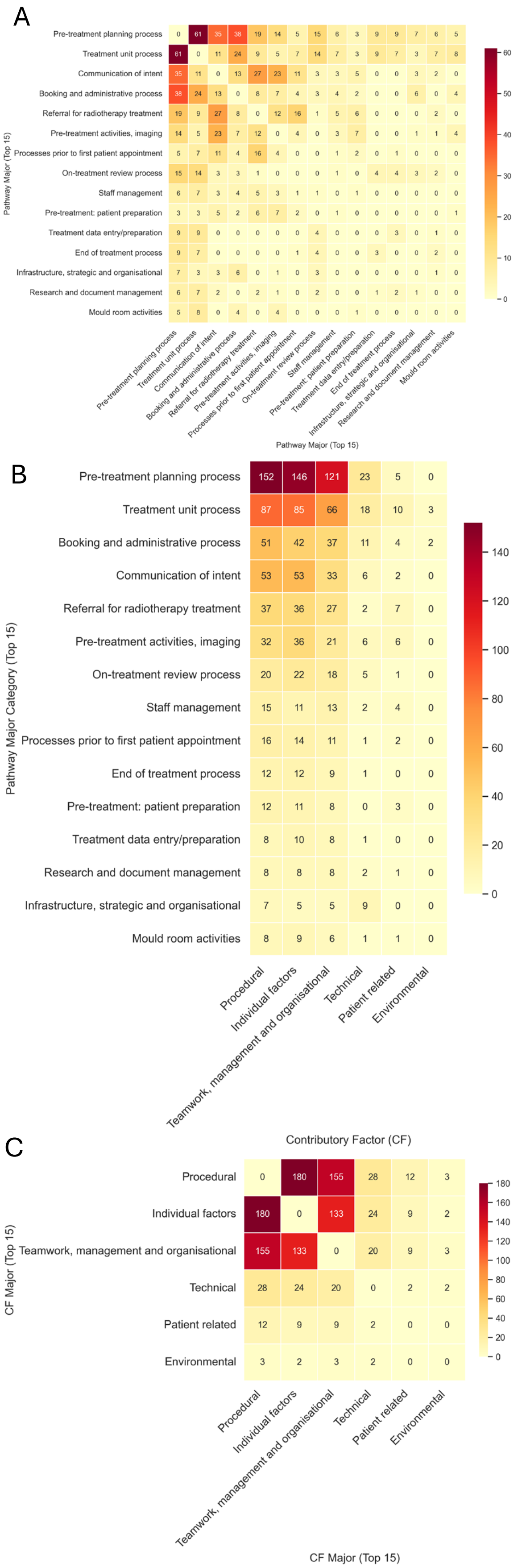
A
Pathway Major (Top 15)
Pre-treatment planning process
Treatment unit process
Communication of intent
Booking and administrative process
Referral for radiotherapy treatment
Pre-treatment activities, imaging
Processes prior to first patient appointment
On-treatment review process
Staff management
Pre-treatment: patient preparation
Treatment data entry/preparation
End of treatment process
Infrastructure, strategic and organisational
Research and document management
Mould room activities
B
Pathway Major Category (Top 15)
Procedural
Individual factors
Teamwork, management and organisational
Technical
Patient related
Environmental
C
Contributory Factor (CF)
CF Major (Top 15)

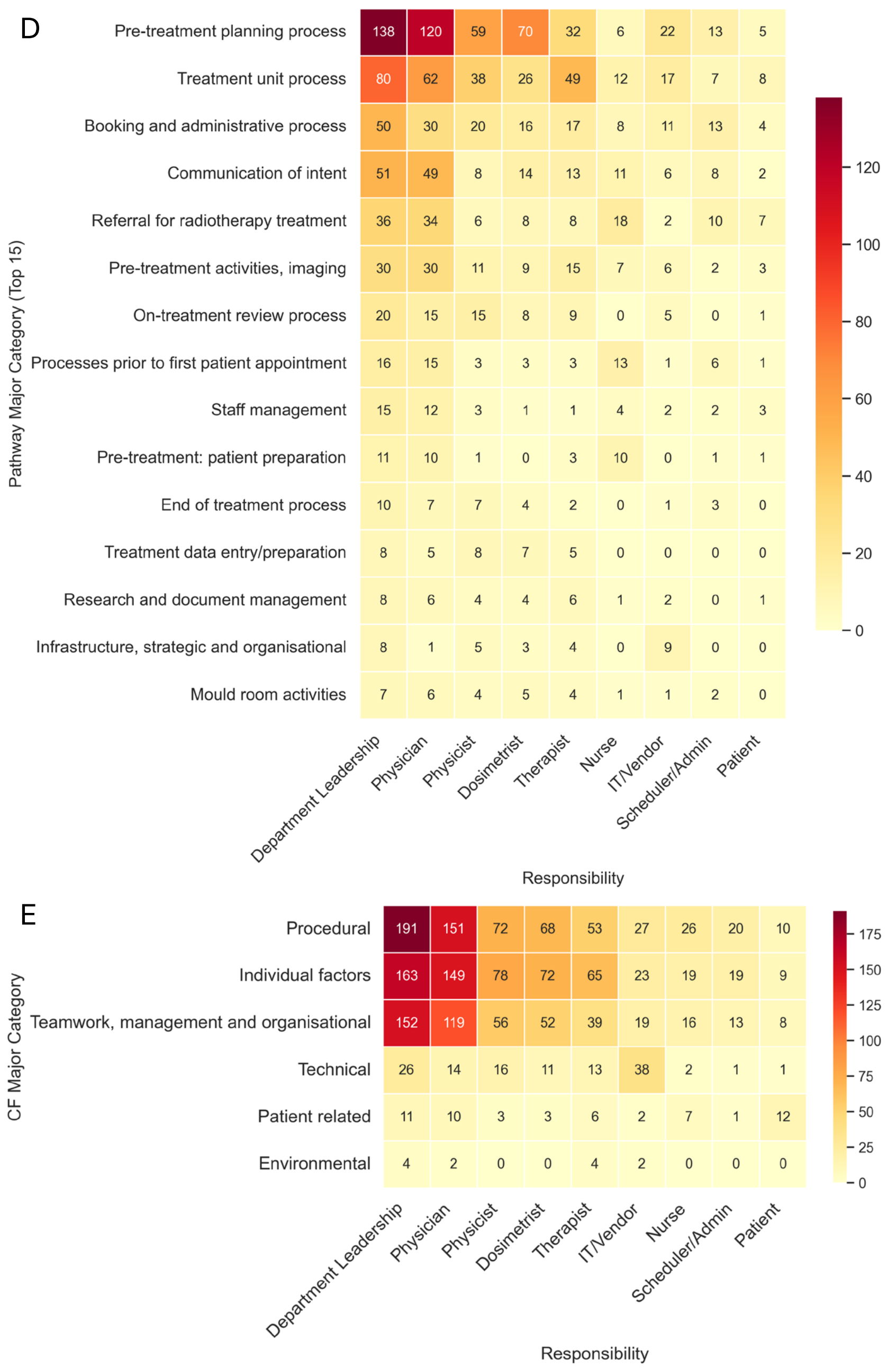

Figure 3 Heatmaps of Co-occurrence Frequencies for Pairwise Combinations of the Five Analyzed Categories. (A) Heatmap between Pathway Categories (Top 15 counts), (B) Heatmap between Pathway (Top 15 counts) and Contributory Factor, (C) Heatmap between Contributory Factor, (D) Heatmap between Pathway (Top 15 counts) and Responsibility, (E) Heatmap between Contributory Factor and Responsibility.

**Drivers of Event Severity**

To identify predictors of incident severity, a series of Ordinal Logistic Regression (OLR) analyses were conducted on the 254 incidents. Separate models were run for radiotherapy pathway steps (Pathway), contributory factors (CF), and professional roles (Responsibility). While the overall explanatory power of these high-level models was limited, they revealed specific significant drivers. The Responsibility model was statistically significant (Pseudo $R^2$ = 0.028, LR $p < 0.001$), as was the Pathway model (Pseudo $R^2$ = 0.033, LR p = 0.015). The Contributory Factor model was not statistically significant overall (Pseudo $R^2$ = 0.010, LR p = 0.059). A detailed analysis of the statistically significant individual predictors ($p < 0.05$) from each model is presented in Table 2.

Table 2 Summary of Ordinal Logistic Regression Analyses Predicting Event Severity

| **Variable Set Analyzed** | **Pseudo $R^2$ (LR p-value)** | **Statistically Significant Predictors ($p < 0.05$)** | **Count (n)** | **Coefficient (β)** | **Odds Ratio (OR)** | **P-value** |
|---|---|---|---|---|---|---|
| Pathway | 0.033 (p = 0.015) | On-treatment review process | 23 | 0.936 | 2.550 | 0.017 |
| | | Research and document management | 8 | 1.612 | 5.013 | 0.047 |
| Contributory Factor | 0.010 (p = 0.059) | Individual factors | 212 | 0.930 | 2.536 | 0.002 |
| Responsibility | 0.028 (p < 0.001) | Therapist | 69 | 1.038 | 2.825 | < 0.001 |

Within the Pathway model, two steps were associated with increased severity. "On-treatment review process," which occurred in 23 incidents, was associated with 2.55 times the odds of a higher severity rating (OR = 2.550, p = 0.017). "Research and document management" (n = 8) showed an even stronger association (OR = 5.013, p = 0.047). For the Contributory Factor model, despite the overall model not reaching statistical significance, the variable "Individual factors" emerged as a highly significant predictor. This factor, present in 212 incidents, was associated with 2.54 times the odds of higher severity (OR = 2.536, $p = 0.002$). Finally, in the Responsibility model, the involvement of a "Therapist" (n = 69) was found to be a highly significant predictor, increasing the odds of a higher severity rating by 2.83 times (OR = 2.825, $p < 0.001$).

**Systemic Vulnerabilities Identified by Association Rule Mining**

To identify systemic vulnerabilities and co-occurring failure modes, Association Rule Mining (ARM) was performed on the dataset of 254 incidents. The analysis was structured to uncover three types of relationships: (1) between different radiotherapy pathway steps (Pathway → Pathway), (2) between pathway steps and contributory factors (Pathway → Contributory Factor), and (3) between different contributory factors (Contributory Factor → Contributory Factor).To ensure the reliability and practical significance of the findings, only high-frequency, high-strength rules were selected for reporting, defined by a minimum co-occurrence Count (n) of 5 and a minimum Confidence of 0.7 (70%).

Analysis of associations between pathway steps identified key process failure clusters (Table 3). Treatment data entry/preparation emerged as a critical node, strongly associated with failures in both Treatment unit process (n = 9, Conf = 0.900) and Pre-treatment planning process (n = 9, Conf = 0.900). Research and document management also showed a strong link to failures at the Treatment unit process (n = 7, Conf = 0.875).

Table 3 Top Association Rules Between Pathways (Count ≥ 5, Confidence ≥ 0.7)

| Antecedent | Consequent | Count | Support | Confidence | Lift |
|---|---|---|---|---|---|
| Treatment data entry/preparation | Treatment unit process | 9 | 0.035 | 0.900 | 2.117 |
| Treatment data entry/preparation | Pre-treatment planning process | 9 | 0.035 | 0.900 | 1.378 |
| Research and document management | Treatment unit process | 7 | 0.027 | 0.875 | 2.058 |
| End of treatment process | Pre-treatment planning process | 9 | 0.035 | 0.750 | 1.148 |
| Research and document management | Pre-treatment planning process | 6 | 0.024 | 0.750 | 1.148 |
| Mould room activities | Treatment unit process | 8 | 0.031 | 0.727 | 1.710 |
| Infrastructure, strategic and organisational | Pre-treatment planning process | 7 | 0.028 | 0.700 | 1.071 |

The analysis linking pathway steps to their underlying causes revealed numerous high-confidence associations (Table 4). A total of 22 distinct rules met the inclusion criteria. Notably, seven of these rules demonstrated 100% confidence, indicating an absolute co-occurrence. For instance, failures in "(19) Research and document management" (n = 8) and "(16) End of treatment process" (n = 12) were, in 100% of cases, associated with both "CF1 Individual factors" and "CF2 Procedural" factors. Similarly, "(20) Staff management" (n = 15) was linked to "CF2 Procedural" with 100% confidence.

The highest-frequency rules underscored the systemic link between workflow processes and human/procedural factors. Failures in "(11) Pre-treatment planning process" were overwhelmingly associated with "CF2 Procedural" factors (n = 152, Conf = 0.916) and "CF1 Individual" factors (n = 146, Conf = 0.880). Other high-count associations included "(6) Booking and administrative process" with "CF2 Procedural" (n = 51, Conf = 0.927) and "(5) Communication of intent" with both CF1 and CF2 (n = 53, Conf = 0.898). A particularly noteworthy finding was the rule linking "(0) Infrastructure, strategic and organisational" to "CF3

Technical”, which, while less frequent (n = 9), showed a very high lift (5.715) and confidence (0.900), suggesting a strong, non-obvious systemic link.

Table 4 Top Association Rules Between Pathways and Contributory Factors (Count ≥ 5, Confidence ≥ 0.7)

| **Antecedent** | **Consequent** | **Count** | **Support** | **Confidence** | **Lift** |
|---|---|---|---|---|---|
| (19) Research and document management | CF5 Teamwork, management and organisational | 8 | 0.031 | 1.000 | 1.608 |
| (12) Treatment data entry/preparation | CF1 Individual factors | 10 | 0.039 | 1.000 | 1.198 |
| (16) End of treatment process | CF1 Individual factors | 12 | 0.047 | 1.000 | 1.198 |
| (16) End of treatment process | CF2 Procedural | 12 | 0.047 | 1.000 | 1.198 |
| (19) Research and document management | CF1 Individual factors | 8 | 0.031 | 1.000 | 1.198 |
| (19) Research and document management | CF2 Procedural | 8 | 0.031 | 1.000 | 1.198 |
| (20) Staff management | CF2 Procedural | 15 | 0.059 | 1.000 | 1.198 |
| (14) On-treatment review process | CF1 Individual factors | 22 | 0.087 | 0.957 | 1.146 |
| (6) Booking and administrative process | CF2 Procedural | 51 | 0.201 | 0.927 | 1.111 |
| (11) Pre-treatment planning process | CF2 Procedural | 152 | 0.598 | 0.916 | 1.097 |
| (0) Infrastructure, strategic and organisational | CF3 Technical | 9 | 0.035 | 0.900 | 5.715 |
| (10) Pre-treatment activities, imaging | CF1 Individual factors | 36 | 0.142 | 0.900 | 1.078 |
| (5) Communication of intent | CF1 Individual factors | 53 | 0.209 | 0.898 | 1.076 |
| (5) Communication of intent | CF2 Procedural | 53 | 0.209 | 0.898 | 1.076 |
| (7) Processes prior to first patient appointment | CF2 Procedural | 16 | 0.063 | 0.889 | 1.065 |
| (11) Pre-treatment planning process | CF1 Individual factors | 146 | 0.575 | 0.880 | 1.054 |
| (14) On-treatment review process | CF2 Procedural | 20 | 0.079 | 0.870 | 1.042 |
| (20) Staff management | CF5 Teamwork, management and organisational | 13 | 0.051 | 0.867 | 1.393 |
| (12) Treatment data entry/preparation | CF5 Teamwork, management and organisational | 8 | 0.031 | 0.800 | 1.286 |
| (14) On-treatment review process | CF5 Teamwork, management and organisational | 18 | 0.071 | 0.783 | 1.258 |
| (16) End of treatment process | CF5 Teamwork, management and organisational | 9 | 0.035 | 0.750 | 1.206 |
| (11) Pre-treatment planning process | CF5 Teamwork, management and organisational | 121 | 0.476 | 0.729 | 1.172 |

Finally, the analysis of inter-factor relationships (Table 5) revealed that contributory factors are highly intertwined. The most prevalent association was "Teamwork, management and organisational" (CF5) leading to "Procedural" (CF2) factors (n = 155, Conf = 0.981). A strong, high-count, and bidirectional relationship was also observed between "Individual factors" (CF1) and "Procedural" (CF2) (n = 180, Conf = 0.849), indicating that these two types of factors co-existed in 70.9% of all incidents.

Table 5 Top Association Rules Between Contributory Factors (Count ≥ 5, Confidence ≥ 0.7)

| **Antecedent** | **Consequent** | **Count** | **Support** | **Confidence** | **Lift** |
|---|---|---|---|---|---|
| Teamwork, management and organisational | Procedural | 155 | 0.610 | 0.981 | 1.175 |
| Individual factors | Procedural | 180 | 0.709 | 0.849 | 1.017 |
| Procedural | Individual factors | 180 | 0.709 | 0.849 | 1.017 |
| Teamwork, management and organisational | Individual factors | 133 | 0.524 | 0.842 | 1.009 |
| Procedural | Teamwork, management and organisational | 155 | 0.610 | 0.731 | 1.175 |

**Statistical Significance of Categorical Associations**

Chi-square tests confirmed that the associations between the different event taxonomies were non-random. As detailed in Table 6, all six pairwise comparisons between the major and sub-level categories for Pathway, Contributory Factors (CF), and Responsibility yielded highly statistically significant results ($p \leq 0.001$).

Table 6 Chi-Square Test Results for Independence Between Event Categories

| Comparison | Chi-square Statistic | P-value | Degrees of Freedom |
|---|---|---|---|
| Pathway vs. CF | 156.718 | 0.001 | 95 |
| Pathway vs. Responsibility | 378.038 | <0.001 | 152 |
| CF vs. Responsibility | 299.845 | <0.001 | 40 |
| Pathway Sub vs. CF Sub | 3642.737 | <0.001 | 2530 |
| Pathway Sub vs. Responsibility | 1355.484 | <0.001 | 848 |
| CF Sub vs. Responsibility | 569.185 | <0.001 | 184 |

ANOVA was conducted to test whether mean event severity scores differed significantly across the groups within each taxonomy (Table 7). The results demonstrate that severity is significantly associated with the Pathway (Major) ($F = 3.129$, $p < 0.001$), Pathway (Sub) ($F = 3.705$, $p < 0.001$), and Responsibility ($F = 2.872$, $p = 0.004$) categories. Critically, this analysis also revealed a key difference based on granularity: while the detailed Contributory Factor (Sub) categories were strongly predictive of severity ($F = 6.835$, $p < 0.001$), the high-level Contributory Factor (Major) categories were not ($F = 1.096$, $p = 0.361$). This finding aligns with the OLR results, reinforcing that detailed, sub-level information is crucial for accurately modeling the drivers of event severity.

Table 7 Analysis of Variance Results Between Event Categories and Severity

| **Analysis** | **F-Statistic** | **P-value** | **Groups Compared** |
|---|---|---|---|
| Pathway vs. Severity | 3.129 | <0.001 | 19 |
| Pathway Sub vs. Severity | 3.705 | <0.001 | 76 |
| CF vs. Severity | 1.096 | <0.361 | 6 |
| CF Sub vs. Severity | 6.835 | <0.001 | 21 |
| Responsibility vs. Severity | 2.872 | 0.004 | 9 |

## Discussion

This study successfully developed and applied an LLM-powered analytical framework to institutional safety incidents, providing a uniquely detailed and quantitative portrait of clinical risk. Our findings demonstrate a clear, data-driven hierarchy of risk drivers and offer a new paradigm for evidence-based safety management in radiation oncology.

OLR analysis of the high-level categories provided insight into the relative predictive power of these broad classifications. Among these models, the Pathway (Major) model demonstrated the strongest relationship with severity (Pseudo $R^2 = 0.033$, LR $p = 0.015$), followed by the Responsibility model (Pseudo $R^2 = 0.028$, LR $p < 0.001$). The Contributory Factor (Major) model showed the weakest relationship and was not statistically significant as a whole (Pseudo $R^2 = 0.010$, LR $p = 0.059$).

A more granular analysis of the individual predictors within these models identified specific, high-leverage targets for intervention. Failures in "Research and document management" (n = 8) were associated with a 5.0-fold increase in the odds of higher severity (OR = 5.013, p = 0.047). Similarly, involvement of a "Therapist" (n = 69) was associated with a 2.8-fold increase (OR = 2.825, p < 0.001), and the presence of "Individual factors" (n = 212) was associated with a 2.5-fold increase (OR = 2.536, p = 0.002). These findings pinpoint specific processes and factors that, regardless of the overall model's fit, carry a disproportionate impact on event severity.

The Association Rule Mining (ARM) analysis complemented these findings by uncovering how failures are interconnected. The analysis highlighted several high-frequency systemic links, such as the association between "(11) Pre-treatment planning process" and "CF2 Procedural" (n = 152, Conf = 0.916), and the strong inter-factor relationship between "Teamwork, management and organisational" and "Procedural" (n = 155, Conf = 0.981). This finding that planning is the most frequent source of error is consistent with previous analyses of institutional and national incident learning systems, which have similarly identified "treatment planning" as the process where incidents most frequently originate[11,12].

The methodology presented in this paper offers a paradigm shift for safety management, moving the field from qualitative, subjective risk assessment toward a quantitative, evidence-based model. Our framework addresses this fundamental limitation directly by proposing a model for an "evidence-based FMEA" where the components of the Risk Priority Number are derived from empirical data. This approach transforms risk assessment from a static, periodic exercise into a dynamic, continuous learning process. As new incident data is collected, the LLM-powered pipeline can automatically process and integrate it, continuously updating the risk models and providing an evolving, near-real-time picture of the department's safety landscape.

The use of an LLM to automate the classification of unstructured incident narratives enables an analysis of unprecedented scale, depth, and objectivity, overcoming the principal barrier that has prevented the widespread use of ILS data for quantitative risk modeling. The multi-modal analytical approach, combining ANOVA, Ordinal Logistic Regression (OLR), ARM, and chi-square tests, provides a holistic and multi-dimensional view of clinical risk that is far richer than what can be achieved with any single method.

However, the study is subject to several limitations. First, the data is derived from a single academic institution, and the findings regarding specific failure modes and their frequencies may not be generalizable to other clinical environments. Second, while the LLM provides a powerful and consistent classification tool, the potential for biases inherent in the LLM's training data exists, although this risk is mitigated by constraining the model's output to established, expert-validated taxonomies. Finally, like all studies based on voluntary incident reporting, this analysis is subject to reporting bias; a well-documented limitation of ILS data wherein near-misses or events of specific types may be under-represented.

The framework established in this study opens several avenues for future research. The most immediate next step is to the types of events that are reported may differ systematically from those that go unreported. This preliminary work guarantees the validate this methodology on a larger, multi-institutional dataset, such as a national RO-ILS database, to test the generalizability of our findings and the scalability of the LLM-powered approach. Further refinement of the LLM prompting and fine-tuning could enhance classification accuracy and allow for the extraction of even more granular data. Ultimately, the analytical engine developed in this study could be integrated into a real-time safety dashboard for clinical leadership, creating a proactive, intelligent, and data-driven safety management system that can help prevent errors before they occur.

## Conclusion

The integration of Large Language Models with incident learning system data represents a transformative step forward for patient safety in radiation oncology. The framework presented in this study demonstrates a feasible and powerful method for moving beyond subjective, prospective risk assessment to an objective, evidence-based, and quantitative model. By systematically converting unstructured narrative data into a rich, structured dataset, this approach enables the identification of the key drivers of event severity and the mapping of complex, systemic failure pathways. Our findings provide a clear, hierarchical understanding of risk-driven first by process, then by system, and least by individual roles, which provides clinical leaders with the specific, data-driven insights needed to design targeted and effective safety interventions. This methodology paves the way for a new generation of intelligent safety systems that can learn from real-world experience to continuously improve the quality and safety of care for all patients undergoing radiation therapy.

**Conflict of Interest Statement for All Authors**

The authors declare no conflict of interest.

**Funding Statement**

The authors declare no funding statement.

**Data Availability Statement for this Work**

Research data are stored in an institutional repository and will be shared upon request to the corresponding author.

**Acknowledgements**

This research was supported Department of Radiation Oncology, University of Miami.